\documentclass[12pt]{article}

\global\arraycolsep=1pt
\usepackage{graphicx}
\usepackage{amssymb}
\usepackage {pstricks}
\newcommand{\beq}{\begin{equation}}
\newcommand{\eeq}{\end{equation}}
\newcommand{\beqa}{\begin{eqnarray}}
\newcommand{\eeqa}{\end{eqnarray}}
\newcommand{\CR}{\nonumber \\}
\newcommand{\m}{\mu}

\renewcommand{\k}{\kappa}

\renewcommand{\theequation}{\thesection.\arabic{equation}}
\renewcommand{\thefootnote}{\fnsymbol{footnote}}

\newcommand{\antisymmetric}{\setlength{\unitlength}{0.5pt}\raisebox{-1.25ex}{\begin{picture}
(10,20)\put(0,20){\line(1,0){10}}\put(0,10){\line(1,0){10}}\put(0,0){\line(1,0){10}}\put(0,0)
{\line(0,1){20}}\put(10,0){\line(0,1){20}}\end{picture}}}

\newcommand{\symmetric}
{\setlength{\unitlength}{0.5pt}
\begin{picture}(20,10)
\put(0,10){\line(1,0){20}}
\put(0,0){\line(1,0){20}}
\put(0,0){\line(0,1){10}}
\put(10,0){\line(0,1){10}}
\put(20,0){\line(0,1){10}}
\end{picture}}

\begin{document}

\begin{titlepage}
\begin{flushright}
{December, 2003}\\
{preprint UT 03-40}\\
{\tt hep-th/0312234} \\
\end{flushright}
\vspace{0.5cm}
\begin{center}
{\Large \bf Geometric transitions, Chern-Simons gauge theory and
Veneziano type amplitudes
}
\vskip1.0cm
{Tohru Eguchi}
\vskip 1.0em
{\it Department of Physics, Faculty of Science \\
University of Tokyo, Tokyo, 113-0033, Japan}
\vskip 0.8cm
{ Hiroaki Kanno}
\vskip 1.0em
{\it 
Graduate School of Mathematics \\
Nagoya University, Nagoya, 464-8602, Japan}
\end{center}
\vskip0.5cm

\begin{abstract}

We consider the geometric transition and compute the all-genus topological 
string amplitudes expressed in terms of Hopf link invariants and topological vertices 
of Chern-Simons gauge theory.
We introduce an operator technique of 2-dimensional CFT 
which greatly simplifies the 
computations. We in particular show that in the case of
local Calabi-Yau manifolds described by toric geometry 
basic amplitudes are written as vacuum expectation values of a product vertex operators 
and thus appear quite similar to the Veneziano amplitudes of the old dual
resonance models. Topological string amplitudes can be easily evaluated using 
vertex operator algebra.

\end{abstract}
\end{titlepage}


\renewcommand{\thefootnote}{\arabic{footnote}}
\setcounter{footnote}{0}

\section{Introduction}

Recently the method of large $N$ duality and the geometric transition has been applied 
to deriving all-genus amplitudes for topological string theory compactified on certain
local Calabi-Yau manifolds \cite{GV,Vafa,AMV,DFG,AKMV}. 
Here one starts from the deformed conifold side and makes use of the
Chern-Simons gauge 
theory and its link invariants and computes the topological string amplitudes of 
the resolved conifold side. When one makes a suitable choice of the non-compact Calabi-Yau 
manifolds, e.g. local ${\bf F}_0$ or its generalizations, one obtains results which  
reproduces the well-known
solutions of the 4-dimensional ${\cal N}=2$ SUSY gauge theories by Seiberg and Witten \cite{SW}.

In a previous article \cite{EK} we have evaluated rigorously
the topological string amplitudes by proving
the propositions proposed by \cite{Iqb,IK-P1,IK-P2} and have shown that the results 
agree exactly 
with the formula of Nekrasov \cite{NekSW} when his parameters $\beta,\hbar$ 
are identified with the 
K\"{a}hler parameters of local Calabi-Yau manifolds. Thus Nekrasov's formula not only reproduces the Seiberg-Witten solution in the 4-dimensional limit $\beta,\hbar\rightarrow 0$ but
encodes the entire information on the number of holomorphic curves of arbitrary degrees 
and genera
in local Calabi-Yau manifolds when $\beta,\hbar\not =0$. (For related more recent works see \cite{Zhou,HIV,Kon}).

In these computations the Hopf link invariants of Chern-Simons theory are expressed in terms of 
Schur functions associated with the representations of Wilson lines and we have to 
evaluate
the sum of the product of link invariants over all representations or Young tableaus. 
We have used certain
identities of (skew) Schur functions in carrying out the computations. It is well-known, 
however, 
Schur functions have a simple 
representation in terms of the vertex operators and there is also a one to one correspondence 
between Young tableaus with the elements of the free fermion Fock space. Thus by making use of 
the operator method of 2-dimensional CFT we can greatly streamline the computations of 
Chern-Simons theory. 

In this article we would like to show that in fact the computation of topological string
amplitude using Chern-Simons theory is very much simplified and 
reduced to evaluating Veneziano-like amplitudes
of the old dual models when the operator method is used; the basic amplitude of the theory 
becomes 
the vacuum expectation value of a product of vertex operators and this observation greatly 
facilitates the calculations.

Such an operator technique has also been discussed recently in detail in the context of the
B-model version of the geometric
transition \cite{ADKMV}. Here our presentation corresponds to the A-model side of the story.

\section{Free fields and the vertex operator}

Let us first introduce a free boson in two dimensions, or the infinite dimensional
Heisenberg algebra with the generators $\alpha_n$ which satisfy the commutation
relation:
\beq
[\alpha_m, \alpha_n] = m \delta_{m, -n}~.
\eeq
Our convention is that $\alpha_n (n>0)$ are the annihilation operators and
$\alpha_{-n} (n>0)$ are the creation operators. (In the following
the zero mode $\alpha_0$ plays no role.) We define the
annihilation $\Gamma_+$ and the creation $\Gamma_-$ part of the vertex operator as follows;
\beq
\Gamma_{\pm}(t_n) := \exp \left( \sum_n t_n \alpha_{\pm n} \right)~.
\eeq
They satisfy
\beq
\Gamma_{\pm}^* (t_n) = \Gamma_{\mp} (t_n)~, \qquad  \Gamma_+ (t_n) |0\rangle = |0\rangle~,
\eeq
and the commutation relation
\beq
\Gamma_+(t_n) \Gamma_-(s_n) = \exp (\sum_n n t_n s_n)  \Gamma_-(s_n) \Gamma_+(t_n)~. \label{Vcomm}
\eeq
We specify the values for the parameters $\{t_n\},\{s_n\}$ soon below.

It is well-known that the
free fermion description of a Young diagram of a representation $R$ is given by 
an element in the free fermion Fock space which is obtained by filling all
the energy levels at
\beq
-(\mu^R_{i} -i+{1\over 2}), \hskip4mm i=1,2,3,\cdots
\eeq 
where $\mu^R_i$ is the length of the i-th row of the Young tableau of
representation $R$.

Let us denote by $\psi^*_n \,(\psi_m)$ the fermion creation (annihilation) 
operators at the level $n\,(m)$. Boson-fermion correspondence is given by the usual formula
\beqa
&&i\partial_z\phi(z)=:\psi(z)^*\psi(z):,\hskip3mm \phi(z)=i\sum_n {1\over n}\alpha_nz^{-n},\\
&&\psi(z)^*=\sum_n\psi^*_nz^{-n-1/2},\hskip3mm \psi(z)=\sum_n\psi_nz^{-n-1/2}.
\eeqa
The Dirac sea corresponding to
the trivial representation $\bullet$ is then given by
\beq
|\mbox{sea}\rangle\equiv|0\rangle=\psi^*_{1/2}\psi^*_{3/2}\psi^*_{5/2}\psi^*_{7/2}\psi^*_{9/2}
\cdots 
|0\rangle\rangle~.
\eeq
If one considers a non-trivial representation, for instance, an
anti-symmetric representation $\!\antisymmetric$~, 
the corresponding state is given by
\beq
|v_{\,\antisymmetric}\rangle=\psi^*_{-1/2}\psi^*_{1/2}\psi^*_{5/2}\psi^*_{7/2}\psi^*_{9/2}\cdots|0\rangle\rangle~.
\label{anti-symm}\eeq 
If instead one considers the symmetric representation $\symmetric$\,, one obtains 
\beq
|v_{\,\symmetric}\rangle=
\psi^*_{-3/2}\psi^*_{3/2}\psi^*_{5/2}\psi^*_{7/2}\psi^*_{9/2}\cdots|0\rangle\rangle~.
\label{symm}\eeq

Now let us set the values of the parameters $\{t_n\}$
of the vertex operator as the power sum of the basic
variables $\{x_i\}$
\beq
t_n={1\over n}p_n={1\over n}\sum_{i=1}x_i^n
\eeq
and introduce the notation
\beq
V_{\pm}(x_i) := \Gamma_{\pm} ( t_n=\frac{1}{n} p_n(x_i))~.
\eeq
Then one defines the Schur functions by the matrix elements
\beq
s_R(x_i)=\langle v_R|V_-(x_i)|0\rangle.
\label{Schur}\eeq
In the case of the anti-symmetric representation of second rank (\ref{anti-symm}), one finds
\beq
s_{\,\antisymmetric}(x_i)=\sum_{i<j}x_ix_j.
\label{Schur for anti-symm}\eeq
Similarly, in the case of the symmetric representation of second rank (\ref{symm}), one has
\beq
s_{\,\symmetric}(x_i)=\sum_i x_i^2+\sum_{i<j}x_ix_j.
\eeq
Thus we obtain elementary (complete) symmetric polynomials of the second order
which in fact agree with the Schur functions associated with these representations.
The formula (\ref{Schur}) gives the general prescription of
representing the Schur functions in terms of vertex operators.

It is easy to generalize (\ref{Schur}) to the case of the skew Schur functions
and represent them
as the matrix elements of the vertex operator
\beq
s_{R/Q}(x) = \langle R | V_-(x_i) | Q \rangle = \langle Q | V_+(x_i) | R \rangle~. \label{vs}
\eeq
Here the Young tableau of the representation $Q$ must be contained in that of $R$.

From the basic relation (\ref{vs}) it is easy to see that the identity for the summation of 
the product of the skew
Schur functions (\ref{Schur1}), for instance, can be reproduced using the commutation 
relation of the vertex
operators. For example,
\beqa
\sum_R s_{R/Q}(x) s_{R/T} (y) &=& \sum_R \langle Q | V_+(x_i) | R \rangle 
\langle T | V_+(y_i) | R \rangle \CR
&=& \langle Q | V_+(x_i) V_-(y_i) | T \rangle \CR
&=& \prod_{i,j \geq 1} ( 1 - x_i y_j)^{-1} \langle Q | V_-(y_i) V_+(x_i) | T \rangle \CR
&=& \prod_{i,j \geq 1} ( 1 - x_i y_j)^{-1} \sum_U \langle Q | V_-(y_i)| U \rangle 
\langle U | V_+(x_i) | T \rangle \CR
&=& \prod_{i,j \geq 1} ( 1 - x_i y_j)^{-1} \sum_U s_{Q/U}(y) s_{T/U}(x)~.
\eeqa
Here we have used
\beq
\prod_{i,j \geq 1} ( 1 - x_i y_j)^{-1}=\exp\left[\sum_{n=1}{1\over n}p_n(x_i)p_n(y_j)\right]~.
\eeq
In the computation of topological string amplitudes, we will need the
"energy" operator $L_0$ appearing in the propagator. In terms of the oscillator 
of the free boson $L_0$ is given by
\beq
L_0 := \sum_{n=1}^\infty \alpha_{-n} \alpha_n~.
\eeq
Note that $L_0$ has no zero mode. The commutation relation
\beq
[L_0, \alpha_n] = -n \alpha_n~,
\eeq
implies that
\beq
Q^{L_0} \alpha_n Q^{-L_0} = Q^{-n} \alpha_n~.
\eeq
Therefore we obtain 
\beq
Q^{L_0} V_{\pm} (x_i) = V_{\pm} ( Q^{\mp 1} x_i) Q^{L_0}~. \label{Qcomm}
\eeq

\section{Sample calculation of Veneziano-like amplitudes}
\setcounter{equation}{0}

The $SU(N)$ supersymmetric gauge theory with eight super charges is
geometrically engineered by local Calabi-Yau manifolds with 
ALE space of $A_{N-1}$ type fibered over ${\bf P}^1$
\cite{KKV, KMV, KV, CKYZ}. The fiber consists of a chain of $(N-1)$ rational curves
which give a minimal resolution of $A_{N-1}$ singularity in (complex) two
dimensions. The dual toric diagram of this local Calabi-Yau
geometry is given by the ladder diagram with $N$ horizontal legs with 4 external lines
as depicted in Fig.1.
By cutting in the middle across $N$ legs as in \cite{IK-P2}, we obtain a pair of
tree diagrams
with $N+2$ legs each. These graphs remind us of the multi-peripheral diagrams of the old dual 
resonance model.
In fact as we will see shortly 
the computation of topological string amplitude using the Hopf link invariant and
topological vertex 
becomes completely analogous to the computation of the Veneziano amplitudes, 
once we use the operator formalism in terms of free fields.


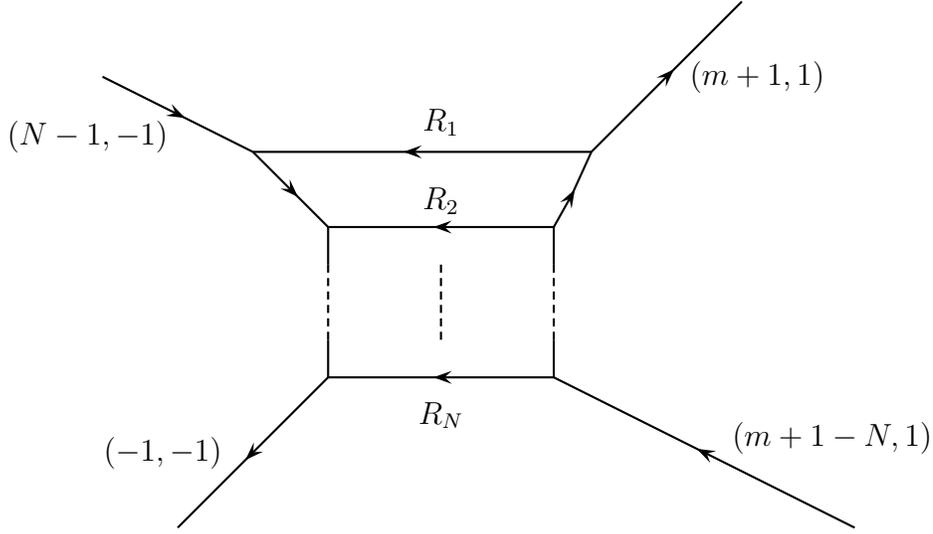
\begin{figure}[h]
\begin{center}
\begin{pspicture}(-3,-3)(8,5)
\psline[arrowsize=5pt]{->}(0,0)(-1.1,-1.1)\psline(-1,-1)(-2,-2)
\psline[arrowsize=5pt]{->}(7,-2)(4.9,-0.95)\psline(5,-1)(3,0)
\psline(0,0)(0,0.5)\psline(0,1.5)(0,2)
\psline[linestyle=dashed,dash=3pt 2pt](0,0.5)(0,1.5)
\psline[linestyle=dashed,dash=3pt 2pt](1.5,0.5)(1.5,1.5)
\psline(3,0)(3,0.5)\psline(3,1.5)(3,2)
\psline[linestyle=dashed,dash=3pt 2pt](3,0.5)(3,1.5)
\psline[arrowsize=5pt]{->}(3,0)(1.4,0)\psline(1.5,0)(0,0)
\psline[arrowsize=5pt]{->}(3,2)(1.4,2)\psline(1.5,2)(0,2)
\psline[arrowsize=5pt]{->}(3.5,3)(1,3)\psline(1.1,3)(-1,3)

\psline[arrowsize=5pt]{->}(-1,3)(-0.4,2.4)\psline(-0.5,2.5)(0,2)
\psline[arrowsize=5pt]{->}(3,2)(3.28,2.5)\psline(3.25,2.45)(3.5,3)
\psline[arrowsize=5pt]{->}(-3,4)(-1.9,3.45)\psline(-2,3.5)(-1,3)
\psline[arrowsize=5pt]{->}(3.5,3)(4.6,4.1)\psline(4.5,4)(5.5,5)

\rput(1.5,3.4){$R_1$}\rput(1.5,2.35){$R_2$}\rput(1.5,-0.5){$R_N$}
\rput(-2.2,-1){$(-1,-1)$}\rput(6.7,-0.8){$(m+1-N,1)$}
\rput(-3.2,3.2){$(N-1,-1)$}\rput(5.7,4){$(m+1,1)$}

\end{pspicture}
\end{center}
\caption{ Ladder diagram for $SU(N)$ gauge theory. 
Note that there are $N+1$ possible toric diagrams ($m= 0, \cdots N$).
$m$ is related to the coefficients of the Chern-Simons coupling in five dimensions.}

\end{figure}


We first recall that the Hopf link invariant $W_{R_1R_2}$ represents the 
expectation value of a pair of linked Wilson lines with 
representations $R_1$ and $R_2$ in Chern-Simons gauge theory and 
is given by the product of Schur functions with the variables $\{x_i\}$ specialized at 
particular values \cite{ML,EK}
\beq
W_{R_1R_2}(q)= s_{R_1}(q^{\rho})s_{R_2}(q^{{\mu_i}^{R_1}+\rho}).
\eeq
Here $q^{\mu+\rho}$ and $q^{\rho}$ means that we specialize the values of $\{x_i\}$ at
\beq
x_i=q^{\mu_i-i+{1\over 2}}\hskip3mm \mbox{and} \hskip3mm x_i=q^{-i+{1\over 2}},\hskip3mm
i=1,2,3,\cdots
\eeq
respectively. On the other hand the topological vertex \cite{AKMV} is defined by 
\beq
C_{R_1, R_2, R_3} 
= q^{\frac{\k_{R_2}}{2}+ \frac{\k_{R_3}}{2}}
s_{R_2^t}(q^\rho) 
\sum_{Q_3} s_{R_1/Q_3}(q^{\m^{R_2^t} + \rho}) s_{R_3^t/ Q_3} ( q^{\m^{R_2} + \rho})~.
\eeq
$R^t$ denotes the conjugate representation of $R$.
$\kappa_R$ and $\ell_R$ of the representation $R$
are defined as usual
\beq
\kappa_R = \ell_R + \sum_{j=1}^{d(R)} \mu_j (\mu_j -2j)~, \quad \ell_R = \sum_{j=1}^{d(R)} 
\mu_j~,
\eeq
where $d(R)$ is the depth of the representation $R$.
Details of the Hopf invariant and topological vertex are relegated to the Appendix.

Let us now compute the amplitudes of the \lq\lq half\rq\rq\ of the $SU(N)$ ladder diagram. See Fig.2.
The value of this amplitude is independent of the parameter $m$ of Fig.1.
We assign to each of the cut $N$ legs associated with the base ${\bf P}^1$ the
representations $R_1, R_2, \cdots R_N$ which are incoming to the vertex by convention.


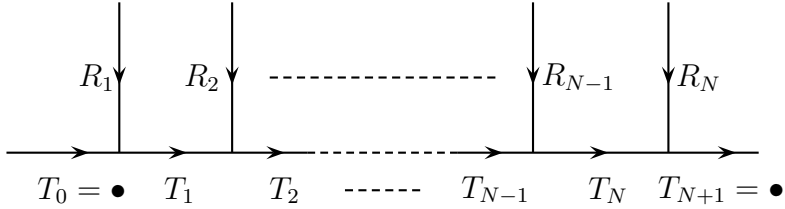
\begin{figure}[h]
\begin{center}
\begin{pspicture}(-3,-1)(8,3) 
\psline[arrowsize=5pt]{->}(-3,0)(-1.9,0)
\psline[arrowsize=5pt]{->}(-2,0)(-0.6,0)
\psline[arrowsize=5pt]{->}(-0.7,0)(0.7,0)
\psline(0.6,0)(1,0)
\psline[linestyle=dashed,dash=3pt 2pt](1,0)(3,0)
\psline[arrowsize=5pt]{->}(3,0)(3.6,0)
\psline[arrowsize=5pt]{->}(3.5,0)(5,0)
\psline[arrowsize=5pt]{->}(4.9,0)(6.7,0)
\psline(6.6,0)(7,0)
\psline[arrowsize=5pt]{->}(-1.5,2)(-1.5,0.9)\psline(-1.5,1)(-1.5,0)
\psline[arrowsize=5pt]{->}(0,2)(0,0.9)\psline(0,1)(0,0)
\psline[arrowsize=5pt]{->}(4,2)(4,0.9)\psline(4,1)(4,0)
\psline[arrowsize=5pt]{->}(5.8,2)(5.8,0.9)\psline(5.8,1)(5.8,0)
\psline[linestyle=dashed,dash=3pt 2pt](0.5,1)(3.5,1)
\rput(-2,-0.5){$T_0=\bullet$}\rput(-0.7,-0.5){$T_1$}\rput(0.7,-0.5){$T_2$}
\psline[linestyle=dashed,dash=3pt 2pt](1.5,-0.5)(2.5,-0.5)
\rput(3.5,-0.5){$T_{N-1}$}\rput(5,-0.5){$T_N$}\rput(6.5,-0.5){$T_{N+1}=\bullet$}
\rput(-1.8,1){$R_1$}\rput(-0.4,1){$R_2$}\rput(4.6,1){$R_{N-1}$}\rput(6.2,1){$R_N$}
\end{pspicture}
\end{center}

\caption{Veneziano-like diagram from cutting in half the $SU(N)$ ladder diagram. 
Note that we respect topology only and ignore the slopes of the edges.}

\end{figure}


To the edges corresponding to the components of the 
fiber we assign representations $T_0, T_1, \cdots T_{N-1}, 
T_N$ with $T_0$ and
$T_N$ being the trivial representations. Then the amplitude is expressed
as the product of $N$ topological vertices of the form
\beq
C_{T_{k-1} R_k T_k^t} \cdot (-1)^{\ell_{T_k}} = (-1)^{\ell_{T_k}}
q^{-\k_{T_k}/2} s_{R_k}(q^{\rho}) \sum_{U_k}  
s_{T_{k-1}/U_k} (q^{\m_{R_k^t} + \rho}) s_{T_{k}/U_k} (q^{\m_{R_k} + \rho})~,
\label{top-vertex1}\eeq
together with the propagators $Q_k^{\ell_{T_k}}  := e^{-{t_k}\cdot{\ell_{T_k}}}$. 
Note that we have an extra sign factor $ (-1)^{\ell_{T_k}}$ 
since $T_k$ is outgoing in our convention.
The parameters $t_k (k=1 \cdots N-1)$ are the K\"ahler
moduli of the $k$-th ${\bf P}^1$ in the chain of rational curves 
and related to the $SU(N)$ gauge theory parameters $a_k$ by
\beq
Q_k = \exp (-2R( a_k - a_{k+1}))~.
\eeq
It is important to take into account the framing factor $(-1)^{n\ell_{T}}q^{-n\kappa_T/2},
\,n\in {\bf Z}$ for
each ${\bf P}^1$ of the fiber. Since the $A_{N-1}$ singularity is fibered over the base, 
each ${\bf P}^1$ of the fiber has 
a zero self-intersection number and the framing index $n$ becomes $n=-1$. 
Thus the factor coming from the framing cancels the factor $(-1)^{\ell_{T_k}}  q^{-\k_{T_k}/2}$ of (\ref{top-vertex1}) and 
we obtain the amplitude
\beq
K^{SU(N)}_{\{R_i\}}=   \sum_{T_1, \cdots, T_{N-1}}  \prod_{k=1}^N Q_k^{\ell_{T_K}} s_{R_k}(q^\rho) 
\sum_{U_k} s_{T_{k-1}/U_k} (q^{\m_{R^t_k} + \rho} ) s_{T_{k}/U_k} (q^{\m_{R_k}+ \rho} )~.
\eeq

We recall that $q^{\m + \rho}$ stands for the substitution $x_i = q^{\m_i -i +1/2}$.
Accordingly let us introduce a notation
\beq
V_\pm^{[R]} (q) := V_\pm ( x_i = q^{\m_i^R -i +1/2})~.
\eeq
We then have
\beqa
& & Q_k^{\ell_{T_k}} \sum_{U_k} s_{T_{k-1}/U_k} (q^{\m_{R_k^t} + \rho} ) s_{T_{k}/U_k} (q^{\m_{R_k}+ \rho} ) \CR
 &=& Q_k^{\ell_{T_k}} \sum_{U_k} \langle T_{k-1} | V_-^{[R_k^t]}(q) | U_k \rangle \langle U_k | 
V_+^{[R_k]}(q) | T_k \rangle \CR
&=&  \langle T_{k-1} | V_-^{[R_k^t]}(q) V_+^{[R_k]}(q) Q_k^{L_0}| T_k \rangle~.
\eeqa
Substituting this, we obtain the following Veneziano-like amplitude for topological string;
\beqa
K^{SU(N)}_{\{R_i\}} 
&=&  \sum_{T_1, \cdots, T_{N-1}} \prod_{k=1}^N s_{R_k}(q^\rho) 
 \langle T_{k-1} | V_-^{[R_k^t]} V_+^{[R_k]} Q_k^{L_0}| T_k \rangle \CR
&=& \prod_{k=1}^N s_{R_k}(q^\rho) \cdot  \langle 0 | \prod_{k=1}^N V_-^{[R_k^t]} V_+^{[R_k]} Q_k^{L_0}| 0\rangle~,\CR
&=&\prod_{i=1}^N {\rm dim}_q\,R_i \cdot \langle 0 | \prod_{k=1}^N V_-^{[R_k^t]} V_+^{[R_k]} Q_k^{L_0}| 0\rangle~.
\label{SU(N)}\eeqa
We have used the fact that $T_0$ and $T_{N+1}$ are the trivial representations and
$s_{R_k}(q^\rho) = {\rm dim}_q~R_k$. (\ref{SU(N)}) is the main result of this paper.

Using the commutation relations (\ref{Vcomm}) and (\ref{Qcomm}) we obtain
\beqa
K^{SU(N)}_{\{R_i\}} (Q_i) &&= \prod_{i=1}^N {\rm dim}_q\,R_i \cdot 
\hskip-5mm \prod_{1\leq m < \ell \leq N}\, \prod_{1\leq i,j} 
 \left( 1 - (\prod_{n=m}^{\ell -1} Q_n) \cdot q^{h_{R_m R_\ell^t} (i,j)} \right)^{-1}~, \CR
&& \hskip-25mm 
= \prod_{i=1}^N {\rm dim}_q\,R_i \cdot \hskip-3mm \prod_{1\leq m < \ell \leq N} \prod_k^\infty
\left( 1 - (\prod_{n=m}^{\ell -1} Q_n) \cdot q^k\right)^{\hskip-1mm -k}\hskip-3mm 
\prod_k \left( 1 - (\prod_{n=m}^{\ell -1} Q_n) \cdot q^k \right)_{~.}^{-C_k(R_m, R_\ell^t)}
\eeqa
$h_{R_1R_2}(i,j)$ denotes the relative hook length (\ref{relative hook})
and 
in deriving the 2nd line we have used the lemma in Appendix.

This is a formula conjectured in \cite{IK-P2} and proved recently in \cite{Zhou}
using Schur function identities. Topological string amplitudes can be immediately constructed 
from the above formula; taking the square of $K_{\{R_i\}}^{SU(N)}$ and multiply the propagators
$e^{-t_B\sum_i l_{R_i}}$ and sum over all representations $\{R_i\}$. One then finds the exact 
agreement with the formula of Nekrasov for $SU(N)$ gauge theory.

\section{Discussions}
\setcounter{equation}{0}

In this paper we have introduced operator techniques in evaluating Chern-Simons amplitudes
in the theory of geometric transition. We find that a free field representation of 
 Schur functions and Young tableaus greatly simplifies the calculation. In the end
 the computation boils down to evaluating Veneziano type amplitudes of old dual resonance
 models. 
 In fact the formula (\ref{SU(N)}) is exactly the form of a vacuum value of a product 
 of vertex operators $V_-^{[R^t]}V_+^{[R]}$: vertex operator carries the quantum number of a 
 representations $R$
 via the specialization of its variables $\{x_i=q^{\mu_i^R-i+1/2}\}$. 
 The amplitudes can then be easily evaluated by using the vertex operator algebra. 
 
 It seems that this is a 
 somewhat simpler structure than suggested by \cite{AKMV} based on the consideration of the
 B-model version of the geometric transition.
 Actually it is rather mysterious why we end up with exactly the free field theory structure in 
 dealing with computation of
 the all-genus topological string amplitudes. A possible answer is that the free field nature
 of the problem originates from the fact that we are dealing with the local Calabi-Yau manifolds where global considerations or constraints are non-existent. It will be very interesting 
 to see how
 far the operator analysis of geometric transition can be applied: cases of compact Calabi-Yau 
 manifolds and also Fano varieties will be the most interesting examples to study.      

\vskip1cm

We would like to thank the participants at the workshop "Quantum Period", IIAS, November
18-19, 2003 for discussions.
H.K. would like to thank H. Awata for helpful discussions.
Research of T.E. and H.K. are supported in part by the Grant-in Aid for Scientific Research (No.15540253 and No.14570073) from Japan Ministry of Education, Culture and Sports.

\setcounter{section}{0}

\section*{Appendix: Hopf link invariants and Topological vertex}
\renewcommand{\theequation}{A.\arabic{equation}}\setcounter{equation}{0}

 The Hopf link invariants $W_{R_1, R_2}(q)$ of $SU(N)$ Chern-Simons gauge 
theory at level $k$ depends on the parameter $q = \exp (\frac{2\pi i}{N+k})$.
It is related to the string coupling $g_s$ via $q=e^{ig_s}$. 
The string coupling  
$g_s$ is the parameter of genus expansion of topological string amplitudes. 
Let $\mu_j$ be the number of boxes in the $j$-th row of
the Young diagram $R$.
We define two integers $\ell_R$ and $\kappa_R$ by
\beq
\ell_R = \sum_{j=1}^{d(R)} \mu_j~, \quad 
\kappa_R = \ell_R + \sum_{j=1}^{d(R)} \mu_j (\mu_j -2j)~,
\eeq
where $d(R)$ is the number of rows of the diagram $R$.
We note that the set of integers $\{ \m_j \}_{j=1}^{d(R)}$ defines a partition of $\ell_R$.
According to \cite{AKMV} the topological vertex is given by 
\beq
C_{R_1, R_2, R_3} = q^{\frac{\k_{R_2}}{2} + \frac{\k_{R_3}}{2}} 
\sum_{Q_1, Q_2} N_{Q_1 Q_2}^{R_1 R_3^t} \frac{W_{R_2^t Q_1} W_{R_2 Q_2}}{W_{R_2}}~,
\eeq
where 
\beq
N_{Q_1 Q_2}^{R_1 R_3^t} = \sum_Q N_{Q Q_1}^{R_1} N_{Q Q_2}^{R_3^t}~,
\eeq
and the tensor $N_{R_1 R_2}^{R_3}$ is the branching coefficient, or the multiplicity of the
representation $R_3$ contained in the tensor product representation $R_1 \otimes R_2$.

In \cite{ML} a formula of the Hopf link invariants $W_{R_1, R_2}$ 
has been obtained. When written in terms of Schur functions it is given by
\beq
W_{R_1 R_2}(q) = s_{R_1}(q^\rho) s_{R_2} (q^{\m^{R_1} + \rho})~. \label{link}
\eeq
where $q^{\m + \rho}\,(q^{\rho})$ means that we make the following substitution;
\beq
s_R (x_i = q^{\m_i - i + \frac{1}{2}})~\hskip4mm (s_R (x_i = q^{- i + \frac{1}{2}}))~.
\eeq
Quantum dimension of the representation $R$ is defined by
\beq
dim_q R\equiv W_R(q)=W_{R\bullet}(q)=s_R(q^{\rho}).
\eeq
An alternate definition of the quantum dimension is
\beq
dim_q R={q^{\kappa_R/4}\over \prod_{(i,j)\in R}\left[h(i,j)\right]}
\label{qdim}\eeq 
where $h(i,j)$ denotes the hook length defined by
\beq
h(i,j)=\mu_i-i+\mu^t_j-j~.
\eeq
By looking at the formula (\ref{qdim}) of $W_R(q)$, we see the following
relation for the Schur polynomials at the specialized value;
\beq
s_{R^t}(q) = q^{-\k_R/2} s_R(q) =  (-1)^{\ell_R} s_R(q^{-1})~.
\eeq

Recall the definition of the skew Schur function
\beq
s_{R_1/R} (x) = \sum_{R_2} N_{R R_2}^{R_1} s_{R_2}(x)~.
\eeq
Hence we have
\beq
C_{R_1, R_2, R_3} 
= q^{\frac{\k_{R_2}}{2}+ \frac{\k_{R_3}}{2}}
s_{R_2^t}(q^\rho) 
\sum_{Q_3} s_{R_1/Q_3}(q^{\m^{R_2^t} + \rho}) s_{R_3^t/ Q_3} ( q^{\m^{R_2} + \rho})~.
\label{topSch}
\eeq
The formula of topological vertex in terms of the special values of the skew
Schur functions was first given in  \cite{ORV}. 
The above expression is slightly different from \cite{ORV}, but this is
more convenient for our purpose.
By taking $R_2 = \bullet$ in (\ref{topSch}) and 
using the cyclic symmetry of $C_{R_1, R_2, R_3}$,  we obtain
\beq
W_{R_1 R_2} 
= q^{\frac{\k_{R_2}}{2}} C_{\bullet R_1 R_2^t} 
= q^{\frac{\k_{R_1}}{2} + \frac{\k_{R_2}}{2}} 
\sum_Q s_{R_2^t/Q}(q^{\rho}) s_{R_1^t/ Q} ( q^{\rho})~. \label{symmetric}
\eeq
Thus we have obtained a manifestly symmetric form of $W_{R_1 R_2}$.
On the other hand, if we take $R_3=\bullet$ in (\ref{topSch}), 
only the trivial representation contributes in the summation over $Q$ and
\beq
W_{R_1 R_2} = q^{\frac{\k_{R_2}}{2}} C_{\bullet R_1 R_2^t} 
= q^{\frac{\k_{R_2}}{2}} s_{R_2^t}(q^{\rho}) s_{R_1}(q^{\m^{R_2+ \rho}})
= s_{R_2}(q^{\rho}) s_{R_1}(q^{\m^{R_2+ \rho}})~.
\eeq
After the exchange of $R_1$ and $R_2$ we find  our original expression of $W_{R_1 R_2}$.

Once the topological vertices are expressed in terms of the (skew) Schur functions,
a summation over representations may be performed by using the following formulas;
\beqa
\sum_R s_{R/R_1}(x) s_{R/R_2}(y) &=& \prod_{i,j \geq 1} (1- x_i y_j)^{-1}
\sum_Q s_{R_2/Q}(x) s_{R_1/Q}(y)~, \label{Schur1} \\
\sum_R s_{R/R_1}(x) s_{R^t/R_2}(y) &=& \prod_{i,j \geq 1} (1+ x_i y_j) 
\sum_Q s_{R_2^t/Q}(x) s_{R_1^t/Q^t}(y)~, \label{Schur2}
\eeqa
For example the formula assumed in \cite{IK-P1} and 
proved as Proposition 1 in \cite{EK} can be derived as follows;
\beqa
K_{R_1 R_2}(Q) &:=& \sum_{R^\prime} Q^{\ell_{R^\prime}} 
W_{R_1 R^\prime}(q) W_{R^\prime R_2}(q) \CR
&=& W_{R_1}(q) W_{R_2}(q) \sum_{R^\prime} 
s_{R^\prime}(Q q^{\m^{R_1} + \rho} ) s_{R^\prime}(q^{\m^{R_2} + \rho}) \CR
&=& W_{R_1}(q) W_{R_2}(q) \prod_{i,j \geq 1} 
\left( 1 - Q q^{h_{R_1 R_2}(i,j)} \right)^{-1}~,
\eeqa
where we have defined the "relative" hook length $h_{R_1 R_2}(i,j)$ by
\beq
h_{R_1 R_2}(i,j) := \m_i^{R_1} -i  +\m_j^{R_2} - j +1~.
\label{relative hook}\eeq
When $R_1=R, R_2 = R^t$ it reduces to the standard hook length.

Let us introduce the following functions
\beqa
&&f_R(q)={q \over (q-1)}\sum_{i\geq1}(q^{\mu_i-i}-q^{-i})~,\\
&&\widetilde{f}_{R_1R_2}(q)={(q-1)^2\over q}f_{R_1}(q)f_{R_2}(q)+f_{R_1}(q)+f_{R_2}(q)+
{q\over (q-1)^2}~.
\eeqa
We have then the following lemma \cite{EK};

{\bf Lemma}
$$
\prod_{i,j \geq 1} \left( 1 - Q q^{h_{R_1 R_2}(i,j)}\right)
= \prod_{k=1}^\infty \left( 1 - Q q^k\right)^k 
\prod_k \left( 1 - Q q^k\right)^{C_k(R_1, R_2)}  \label{lemma}~,
$$
where $C_k(R_1,R_2)$ are the expansion coefficients of $\widetilde{f}_{R_1R_2}(q)$
\beq
\widetilde{f}_{R_1R_2}(q)
=\sum_k C_k(R_1, R_2) q^k + \frac{q}{(q - 1)^2}~.
\eeq
By this lemma we find that
\beq
\prod_{i,j \geq 1} 
\left( 1 - Q q^{h_{R_1 R_2} (i,j) }\right)^{-1}
= \exp \left( \sum_{n=1}^\infty \frac{Q^n}{n} \frac{q^n}{(q^n -1)^2} \right)
\cdot \prod_k \left( 1 - Q q^{k} \right)^{-C_k(R_1, R_2)}~,
\eeq
which is Proposition 1 in \cite{EK}.



\begin{thebibliography}{99}


\bibitem{GV} R.~Gopakumar and C.~Vafa,
On the Gauge Theory/Geometry Correspondence,
{\it Adv. Theor. Math. Phys.} {\bf 3} (1999) 1415,
{\tt hep-th/9811131}.


\bibitem{Vafa} C.~Vafa,
Superstrings and Topological Strings at large $N$,
{\it J.Math.Phys.} {\bf 42} (2001) 2798,
{\tt hep-th/0008142}.


\bibitem{AMV} M.~Aganagic,  M.~Mari\~no  and C.~Vafa, 
All Loop Topological String Amplitudes From Chern-Simons Theory,
{\tt hep-th/0206164}.

\bibitem{DFG} D.E.~Diaconescu, B.~Florea and A.~Grassi,
Geometric Transitions and Open String Instantons, 
{\it Adv.Theor.Math.Phys.} {\bf  6} (2003) 619,
{\tt hep-th/0205234}; 
Geometric Transitions, del Pezzo surfaces and Open String Instantons,
{\it Adv.Theor.Math.Phys.} {\bf 6} (2003) 643,
{\tt hep-th/0206163}.

\bibitem{AKMV} M.~Aganagic,  A.~Klemm, M.~Mari\~no 
and C.~Vafa, The Topological Vertex,
{\tt hep-th/0305132}.

\bibitem{SW} N.~Seiberg and E.~Witten,
Electric-Magnetic Duality, Monopole Condensation, 
and Confinement in $N=2$ Supersymmetric Yang-Mills 
Theory, {\it Nucl. Phys. } {\bf B426} (1994) 19, 
{\tt hep-th/9407087};
Monopole, Duality, and Chiral Symmetry Breaking 
in $N=2$ Supersymmetric QCD, 
{\it Nucl. Phys. } {\bf B431} (1994) 484, 
{\tt hep-th/9408099}.


\bibitem{EK} T.~Eguchi and H.~Kanno,
Topological Strings and Nekrasov's Formulas,
{\tt hep-th/0310235}.

\bibitem{Iqb} A.~Iqbal, All Genus Topological Amplitudes and 
5-brane Webs as Feynman Diagrams, {\tt hep-th/0207114}.

\bibitem{IK-P1} A.~Iqbal and A.-K. Kashani-Poor,
Instanton Counting and Chern-Simons Theory, {\tt hep-th/0212279}.

\bibitem{IK-P2} A.~Iqbal and A.-K. Kashani-Poor,
$SU(N)$ Geometries and Topological String Amplitudes, {\tt hep-th/0306032}.

\bibitem{NekSW} N.~Nekrasov, Seiberg-Witten Prepotential
from Instanton Counting, {\tt hep-th/0206161}.


\bibitem{Zhou} Jian Zhou,
Curve Counting and Instanton Counting,
{\tt math.AG/0311237}.


\bibitem{HIV} T.~Hollowood, A.~Iqbal and C.~Vafa,
Matrix Models, Geometric Engineering and Elliptic Genera,
{\tt hep-th/0310272}.


\bibitem{Kon} 
Y.~Konishi and K.~Sakai, Asymptotic Form of Gopakumar-Vafa Invariants
from Instanton Counting, {\tt hep-th/0311220};
Y.~Konishi,
Topological Strings, Instantons and Asymptotic Forms of Gopakumar-Vafa Invariants,
 {\tt hep-th/0312090}.

\bibitem{ADKMV} M.~Aganagic, R.~Dijkgraaf, A.~Klemm, M.~Marino and C.~Vafa,
Topological Strings and Integrable Hierarchies,
{\tt hep-th/0312085}.

\bibitem{KKV} S.~Katz, A.~Klemm and C.~Vafa,
Geometric Engineering of Quantum Field Theory,
{\it Nucl. Phys.} {\bf B 497} (1997) 173,
{\tt hep-th/9609239}.

\bibitem{KMV} S.~Katz,  P.~Mayr and C.~Vafa,
Mirror Symmetry and Exact Solution of 4D $N=2$ Gauge Theories I,
{\it Adv. Theor. Math. Phys.} {\bf 1} (1998) 53,
{\tt hep-th/9706110}.

\bibitem{KV} S.~Katz and C.~Vafa,
Matter from Geometry,
{\it Nucl. Phys.} {\bf B497} (1997) 146,
{\tt hep-th/9606086}.

\bibitem{CKYZ} T.-M. Chiang, A.~Klemm, S.-T. Yau and E.~Zaslow, 
Local Mirror Symmetry: Calculation and Interpretation,
{\it Adv. Theor. Math. Phys.} {\bf 3} (1999) 495, {\tt hep-th/9903053}.




\bibitem{ML} H.R~Morton and S.G~Lukac, 
The HOMFLY polynomial of the Decorated Hopf Link,
{\tt math.GT/0108011}.

\bibitem{ORV} A.~Okounkov, N.~Reshetikhin and C.~Vafa,
Quantum Calabi-Yau and Classical Crystals, {\tt hep-th/0309208}.

\end{thebibliography}
\end{document}